\documentclass[12pt]{article}

\def\textit#1{{\color{blue} #1}}

\usepackage[body={17.5cm, 22.5cm},right=2cm]{geometry}

\usepackage{float} 

\usepackage{booktabs}
\usepackage{caption}

\usepackage{color}
\usepackage{graphicx}
\usepackage{epsf}
\usepackage{graphicx,epsfig}
\usepackage{amsmath}

\usepackage{multirow}

\usepackage{amsmath, amsthm, amssymb}

\usepackage{epsfig}
\usepackage{cite}
\usepackage{color,colordvi}

\newcounter{hran}

\makeatletter
\renewcommand\section{\@startsection {section}{1}{\z@}%
                               {-3.5ex \@plus -1ex \@minus -.2ex}%
                               {2.3ex \@plus.2ex}%
                               {\normalfont\large\bfseries}}
\makeatother

\begin{document}\thispagestyle{empty}

\vspace{0.5cm}

\def\thefootnote{\arabic{footnote}}
\setcounter{footnote}{0}

\def\s{\sigma}
\def\nn{\nonumber}
\def\p{\partial}
\def\ls{\left[}
\def\rs{\right]}
\def\lc{\left\{}
\def\rc{\right\}}
\def\S{\Sigma}
\def\l{\lambda}
\newcommand{\beq}{\begin{equation}}
\newcommand{\eeq}[1]{\label{#1}\end{equation}}
\newcommand{\bea}{\begin{eqnarray}}
\newcommand{\eea}[1]{\label{#1}\end{eqnarray}}

\newcommand{\be}{\begin{eqnarray}}
\newcommand{\ee}{\end{eqnarray}}
\newcommand{\bi}{\begin{itemize}}
\newcommand{\ei}{\end{itemize}}
\renewcommand{\th}{\theta}
\newcommand{\bth}{\overline{\theta}}

\hspace*{12cm}
CERN-PH-TH/2013-216

\vspace*{2cm}

\begin{center}

{\Large \bf 
Supersymmetry Breaking by Higher Dimension Operators
 }
\\[1.5cm]
{\normalsize \bf Fotis Farakos$^{1}$,\,  Sergio Ferrara$^{2,3,*}$,  Alex Kehagias$^{1,4}$ and Massimo Porrati$^{5}$
}
\\[0.5cm]

\vspace{.3cm}
{\small { $^{1}$ Physics Division, National Technical University of Athens, 15780 Zografou Campus, 
Athens, Greece}}\\

\vspace{.1cm}
{\small {  $^{2}$ Physics Department, Theory Unit, CERN,
CH 1211, Geneva 23, Switzerland}}\\

\vspace{.1cm}
{\small {  $^{3}$ INFN - Laboratori Nazionali di Frascati,
Via Enrico Fermi 40, I-00044 Frascati, Italy}}\\

\vspace{.1cm}
{\small {  $^{4}$ Department of Theoretical Physics
24 quai E. Ansermet, CH-1211 Geneva 4, Switzerland}}\\

\vspace{.1cm}
{\small {  $^{5}$
CCPP, Department of Physics, NYU 4 Washington Pl. New York NY 10016, USA}}



\end{center}

\vspace{1cm}

\begin{center}
{\small  \noindent \textbf{Abstract}} \\[0.5cm]
\end{center}
\noindent 
{\small We discuss a  supersymmetry breaking mechanism 
for  ${\cal{N}}=1$ theories triggered by higher dimensional operators. 
We consider  such operators for real linear and chiral spinor superfields
that break superymmetry and reduce to the Volkov-Akulov action.
We also consider  supersymmetry breaking induced by a higher dimensional operator of a nonminimal scalar (complex linear) multiplet.
The latter  differs from the standard chiral multiplet in its auxiliary sector, which 
contains,  in addition to the complex scalar auxiliary of a chiral  
superfield, a complex vector and two spinors auxiliaries. 
By adding an appropriate higher dimension operator, the scalar auxiliary may   
acquire a nonzero {\rm {vev}} triggering spontaneous supersymmetry breaking. We find that the spectrum of the theory in the 
supersymmetry breaking vacuum consists of a  free chiral multiplet and a constraint chiral superfield describing
the goldstino. Interestingly, the latter  turns out to be one 
of the auxiliary fermions, which becomes dynamical in the supersymmetry breaking vacuum. In all 
cases we are considering here, there is no sgoldstino mode and thus the goldstino does not have a superpartner. 
The sgoldstino is decoupled since the goldstino is one of the auxiliaries, which  is propagating 
only in the supersymmetry breaking vacuum.
We also point out how  higher dimension operators introduce a potential for the propagating scalar of the theory.

}
\vskip 1cm

\def\thefootnote{\arabic{footnote}}
\setcounter{footnote}{0}

\vskip.5in
\line(1,0){250}\\
{\footnotesize {$^*$On leave of absence from Department of Physics and Astronomy, University of California Los Angeles, CA 90095-1547
USA}}




\baselineskip= 19pt
\newpage 

\section{Introduction}

Supersymmetry  is one of the most appealing candidates for new physics. It
has not been observed so far; thus, it should be broken at some high energy scale if it is realised at all.
The central role on how supersymmetry is broken is usually played by the scalar potential of the supersymmetry
breaking sector. Scalar potentials in supersymmetry and supergravity have  extensively been studied for
two-derivative theories. Even though it is known that introducing higher dimension operators 
spoils the form of the scalar potential, it seems that the  theory somehow protects itself  from unconventional
non-supersymmetric vacua \cite{fer}. Our task here is to discuss how scalar potentials are modified and may lead to supersymmetry breaking when
higher dimension operators are introduced. 
The goldstone fermion associated with the supersymmetry breaking, the goldstino, is
described by the Volkov-Akulov action~\cite{Akulov}, in which supersymmetry is non-linearly realized. In particular, the goldstino dynamics has  been 
related in \cite{ks} to the
superconformal anomaly multiplet $X$ corresponding to the FZ supercurrent \cite{fz}. 
The multiplet of anomalies X, defined in the UV
flows in the IR, under renormalization group, to a chiral superfield $X_{NL}$ which obeys  the constraint 
$X_{NL}^2=0$. 
This constrained superfield is the realization of the goldstino given in~\cite{feruglio}. 
Since the dynamics of the goldstino is universal, the IR action in \cite{ks} is the same as in~\cite{feruglio}. 
Constrained superfields have been used before to accomodate the goldstino. Indeed, 
there are alternative formulations in which the goldstino sits in a constrained superfield, such as a constrained
chiral  multiplet~\cite{rocek}, a constrained vector multiplet~\cite{rocek2}, a spinor superfield~\cite{wess}, or a complex linear
superfield~\cite{kuzenko}. Constrained superfields have also been used recently in the MSSM context
\cite{ignatios,brignole,petersson,far0} and in inflationary cosmology, where the inflaton is 
identified with the sgoldstino
\cite{luis}. In addition their interaction with matter has been worked out in \cite{ignatios1}.

Supersymmetric theories that contains higher dimension operators (derivative or non-derivative ones)
have some novel features 
\cite{fer1,ovrut,Far1,Far2}.
Among these, an interesting aspect  is that higher dimension operators can
contribute  to the scalar potential. This has been discussed earlier in \cite{fer} where a few examples have been given.  
In particular, theories with no potential at the leading two-derivative level, 
may develop a nontrivial potential when higher dimension operators are taken into
account and may even lead to supersymmetry breaking, as already mentioned above. 
At this point there are however, two dangerous aspects. The first one concerns
the appearance of ghost instabilities. In the type of theories we are discussing,
this instability is not present as the theory does not have those  higher derivatives terms which might give 
rise to such dangerous states.
The second issue concerns the auxiliary
fields. Here, we are still able to eliminate the auxiliaries of the  multiplet since they
appeared algebraically in the supersymmetric Lagrangian.

We will consider various theories exhibiting supersymmetry 
breaking in the presence of higher dimension operators. 
Special attention will be devoted to
a globally supersymmetric model for a complex linear multiplet. As we will
explain in one of the following sections, the complex linear multiplet, or nonminimal multiplet, contains the degrees 
of freedom of a chiral multiplet and in addition, two  fermions and a complex vector. At the two derivative level,
 both the extra fermions and the complex vector are auxiliaries and can be integrated out, giving on-shell
just a free complex scalar and a fermion. Due to the constraints
the complex linear 
satisfies, there is no superpotential one can write down and  the introduction
of an F-term for non-derivative interactions is not possible. So, one relies on modifying the D-term in order to get some 
non-trivial interactions and an emerging potential induced by higher dimension operators
\cite{fer,ovrut,Far1,Far2}. Under certain conditions, it may happen that the new potential 
develops another extremum for the auxiliaries which break supersymmetry. In this case, new phases will emerge, only 
one of which 
will be realized when the higher dimension operators interactions are turned off. 
It should be noted however, that these new phases are not different phases of the same theory, but rather different
theories.  The examples studied in~\cite{fer} were
not successful in this respect, basically because the auxiliaries appeared in the higher derivative terms with the same sign as in the leading two-derivative term. 
This has the effect that the minimum of the potential is stable with respect to 
the 
addition of the higher dimension term. 
However, in the case of the complex
linear multiplet, the auxiliary in the two derivative term and in the higher derivative term appear with opposite sign. This has the effect of introducing 
now a new minimum for a non zero value of the auxiliary, thereby breaking supersymmetry. The interesting phenomenon 
that appears here is that the  goldstino turns out to be one of the auxiliary fermions of the multiplet, which  
in the new vacuum acquires a kinetic term, but vanishes in the supersymmetric vacuum of the theory. After
integrating out the auxiliaries, we are left with a complex scalar, a fermion and a goldstino without supersymmetric
partner, as supersymmetry is broken. Therefore, there is a mismatch  of bosonic and fermion degrees of freedom as for 
example in Volkov-Akulov type of models where supersymmetry is non-linearly realised \cite{Akulov}. 

This paper is organized as follows. In the next section we present 
theories with higher dimensional operators that exhibit susy breaking and the corresponding Volkov-Akulov actions. 
In section 3 we describe the complex linear multiplet. 
In section 4 we show how higher dimensional operators of  the complex linear multiplet 
may lead to susy breaking and we prove the equivalence to non-linear realizations. 
Finally, we conclude in the last section 5.

\section{SUSY Breaking and Volkov-Akulov Actions}

One of the  explicit examples considered in \cite{fer} to demonstrate that the scalar potential is sensitive to the 
addition of higher dimension terms, is a supersymmetric $\sigma$-model with four-derivative coupling. 
Its standard  Lagrangian is\footnote{Our superspace conventions can be found in \cite{W&B}.}
\be
{\cal L}_{\sigma}=\int d^4 \theta K(\Phi,\bar \Phi),
\ee
where $K(\Phi,\bar \Phi)$ is the K\"ahler potential. 
The latter can be considered as a composite vector multiplet possessing 
an effective gauge (K\"ahler) invariance
\be
K\to K+i(\Lambda-\bar \Lambda),
\ee
where $\Lambda$ is a chiral superfield. As we are going to keep this
invariance for the higher dimension operators as well, we will 
construct the latter in terms of the superfield field
strength 
\be
W_\alpha=-\frac{1}{4}\bar D \bar D D_\alpha K
\ee
for the composite vector $K(\Phi,\bar \Phi)$. Then, clearly, the most general K\"ahler invariant Lagrangian up to
four-derivative terms is
\be
{\cal L}_{\sigma}=\int d^4 \theta\,  K(\Phi,\bar \Phi) +
\left(\int d^2 \theta \, g(\Phi)+\lambda \int d^2 \theta\,  W^2(K)+h.c.,
\right) \label{f1}
\ee
where $g(\Phi)$ is the superpotential and $\lambda>0$. Without loss of generality, let us consider the simplest 
case of a single chiral multiplet with $K=\Phi\bar \Phi$ and $g(\Phi)=0$. Then  
eq.~(\ref{f1}) turns out to be
\be
{\cal L}_{\sigma}=\int d^4 \theta\, \left( \Phi\bar \Phi 
+\frac{\lambda}{2} D^\alpha \Phi D_\alpha \Phi
 \bar D _{\dot{\alpha}}\bar \Phi \bar D ^{\dot{\alpha}} \bar \Phi 
\right) \label{f11}
\ee
and the scalar potential turns out to be \cite{fer}
\be
-V_F= |F|^2+ 8\lambda |F|^4 \label{vf}.
\ee
The minimum of the potential is at $F=0$, which is also the minimum of the theory in the $\lambda\to 0$ limit. 

\subsection{Chiral Spinor Superfield}
There are other possibilities one may wish to consider. For example, let us consider the Lagrangian~[cfr.~\cite{ks,feruglio}]
\be
{\cal L}_W=\frac{1}{4}\left( \int d^2 \theta \, W^\alpha W_\alpha +h.c\right)+\frac{1}{\Lambda^4}
\int d^4\theta\,  W^\alpha W_\alpha \bar W _{\dot{\alpha}} \bar W ^{\dot{\alpha}} \label{ww},
\ee
where 
\be
W_\alpha&=&\lambda_\alpha + \theta_\alpha D + \theta^\beta F_{\alpha\beta} + \theta^2 \chi_\alpha,
\ee
so that $W_\alpha$ is chiral but otherwise unconstrained and $F_{\alpha\beta} = F_{\beta \alpha}$.

The component form of the  Lagrangian (\ref{ww})  is
\be
\nn
{\cal L}_W &=&\frac{1}{4} (D^2 + 2 \chi \lambda +  \frac{1}{2} F^{\alpha \beta}F_{\alpha \beta}  +h.c.) 
\\
\nn
&+&\frac{1}{\Lambda^4} [\l^2 \p^2 \bar \l ^2 
+ (D^2 + 2 \chi \lambda +  \frac{1}{2} F^2)(\bar D^2 + 2\bar \chi \bar \lambda +  \frac{1}{2} \bar F^2)]
\\
\label{fullcomp}
&-& i \frac{1}{\Lambda^4} (\l^{\alpha} D - F^{\alpha \beta} \l_{\beta}  ) 
\s^{\mu}_{\alpha \dot{\alpha}} \p_\mu (\bar \l^{\dot{\alpha}}\bar D - \bar F^{\alpha \beta}\bar \l_{\dot{\beta}} )
\ee
where 
\be
F^{\alpha \beta} = \epsilon^{\alpha \s} \epsilon^{ \beta \rho} F_{\s \rho} ,
\ee

In the particular case that $W_\alpha$
is the field-strength superfield
 and satisfies $D^\alpha W_\alpha=\bar D _{\dot{\alpha}} \bar W ^{\dot{\alpha}}$, the Lagrangian has been 
worked out in \cite{fer,Far2}. The Lagrangian (\ref{ww}) is of the form~\cite{ks,feruglio}
\be
{\cal L}_W=\int d^4 \theta \, X\bar X+\frac{\Lambda^4}{4}
\left(\int d^2\theta\, X+h.c\right) \label{xw}
\ee 
where  $X=W^\alpha W_\alpha$ satisfies 
\be
X^2=0\, .
\ee
The explicit form of $X$ is
\bea
X&=& W^\alpha W_\alpha
= \lambda^2 
+2 \theta^\beta ( \epsilon_{ \beta \alpha} D - F_{ \beta\alpha} ) \lambda^\alpha
+ ( \frac{1}{2} F^{\alpha \beta}F_{\alpha \beta} + D^2 + 2\chi\lambda) \theta^2
\eea{sm1}
with $F^{\alpha \beta} = \epsilon^{\alpha \rho} \epsilon^{ \beta \sigma} F_{\rho \sigma} $.
By defining 
 \be 
 G_\beta= 2  \lambda_\beta  D - 2 F_{ \beta\alpha}  \lambda^\alpha
 \ee
 and noticing  that, because $\lambda^2 \lambda _\alpha =0$,
 
 \be
G^2=\lambda^2( 4 D^2 + 2 F^{\alpha \beta}F_{\alpha \beta} ) 
= \lambda^2( 4 D^2 + 2 F^{\alpha \beta}F_{\alpha \beta}+ 8\chi\lambda  ) \equiv 4 \lambda^2 {\cal F},
\ee
we get the parametrization of $X$ in chiral coordinates~\cite{feruglio,ks}
\beq
X= \frac{ {\tilde G}^2}{ 2{  \cal F}} + \sqrt{2} \theta {\tilde G} + \theta^2 {\cal F}.
\eeq{sm2}
Here we have rescaled $G = \sqrt{2} {\tilde G}$.
In a sense, $W_\alpha$ is the square root of the goldstino. 
If the above form of $X$ 
is plugged back in eq.~(\ref{xw}),  the Volkov-Akulov Lagrangian for the goldstino $G$ is obtained~\cite{ks,feruglio}.

We should note here that the resulting Lagrangian is written entirely in terms of the goldstino $G_\alpha$. 
One would expect the theory to propagate also its supersymmetric partner, the sgoldstino
to fill together  a multiplet of the (broken) susy. However, it seems that the sgoldstino has been integrated out from the 
theory. This is due to the fact that the original multiplet didn't have any propagating fields as both fermions
$\chi,\lambda$ and bosons $D,F_{\alpha \beta}$ were auxiliaries. In a sense, the original theory can be considered as 
the zero-momentum limit (or infinite mass limit) 
of a theory were all fields were propagating. This is equivalent to 
sgoldstino decoupling \cite{feruglio,ks,ignatios,ignatios1,far0} and we correctly find here
that the goldstino is the only propagating mode in the susy broken branch.

A way to find the {\rm vev} of ${\cal F}$
is from the bosonic part of (\ref{ww}), which 
turns out to be
\begin{align}
{\cal L}^B_W&=\left(\frac{1}{8}F^{\alpha\beta}F_{\alpha\beta}
+\frac{1}{4}D^2+h.c\right)+\frac{1}{\Lambda^4}
\left(D^2+\frac{1}{2}F^{\alpha\beta}F_{\alpha\beta}\right)\left(\bar D ^2+\frac{1}{2}\bar F ^{\dot{\alpha} \dot{\beta}}
\bar F _{\dot{\alpha} \dot{\beta}}\right).
\end{align}
The are now two solutions for $D$, 
\be
&&i)\, ~D=0\, , \\
&&ii)\, ~D^2=-\frac{1}{2}F^{\alpha\beta}F_{\alpha\beta}-\frac{\Lambda^4}{4} , ~~~
\bar D ^2=-\frac{1}{2}\bar F ^{\dot{\alpha} \dot{\beta}}\bar F _{\dot{\alpha} \dot{\beta}}-\frac{\Lambda^4}{4}.\label{susy5}
\ee
The first solution 
is  the supersymmetric Lorentz-invariant vacuum,
provided $F_{\alpha\beta}=0$, whereas the second solution gives
\be
{\cal F}=-\frac{\Lambda^4}{4}.
\ee
 Then $\langle F_{\alpha\beta}\rangle \neq 0$  clearly breaks supersymmetry but also Lorentz invariance
 at the same time. 
 However, it is possible to preserve  Lorentz invariance  if   
$\langle F_{\alpha\beta}\rangle =0$
and  $\langle F^{\alpha \beta}F_{\alpha \beta}\rangle\neq 0$ as required by (\ref{susy5}).

In the particular case in which $W_\alpha$ is the field strength superfield, the bosonic part of (\ref{ww}) turns out to be
\cite{Far2}
\begin{align}
{\cal L}^B_W&=-\frac{1}{4}F^{\mu\nu}F_{\mu\nu}-\frac{i}{8}\epsilon^{\mu\nu\kappa\lambda} F_{\mu\nu}F_{\kappa\lambda}
+\frac{1}{2}D^2\nonumber \\
&+\frac{1}{\Lambda^4}
\left\{ 
\frac{1}{4}(F^{\mu\nu}F_{\mu\nu})^2-F^{\mu\nu}F_{\mu\nu}D^2+\frac{1}{16}
(\epsilon^{\mu\nu\kappa\lambda} F_{\mu\nu}F_{\kappa\lambda})^2+D^4\right\}.
\end{align}
There are two solutions for $D$, 
\be
&&i)\, ~D=0\, , \\
&&ii)\, ~D^2=\frac{1}{2}F^{\mu\nu}F_{\mu\nu}-\frac{\Lambda^4}{4}. \label{susy41}
\ee
The first solution corresponds to the supersymmetric branch, 
whereas the second solution gives the possibility $<D^2> \ne 0$ and may break supersymmetry. 
However, this is not a Lorentz-invariant vacuum, since (\ref{susy41}) requires
a non-vanishing $F^{\mu\nu}F_{\mu\nu}$ for supsersymmetry breaking. 
In particular, since $D^2$ is positive, this vacuum can only be sustained 
with a non-zero background magnetic field.

\subsection{Real Linear Multiplet}

Another interesting example is provided by the Lagrangian
\beq
L=\int d^4 \theta \left( -L^2  + {1\over 64\Lambda^4} {D}^{\alpha} L {D}_{\alpha} L \bar D_{\dot\alpha }L 
\bar D^{\dot\alpha}L  \right),
\eeq{m3}
where $L$ is a real linear multiplet. 
The grassmann expansion of the latter may be written as 
\be
\label{sm3}
L =\phi + \theta \psi + \bar\theta \bar \psi - \theta \sigma_\mu \bar \theta H^\mu 
-  {i \over 2 } \theta^2 \bar\theta   \bar \sigma ^\mu \p_\mu \psi  
+ {i \over 2 } \bar\theta^2 \theta \sigma^\mu \p_\mu \bar\psi 
- {1 \over 4 } \theta^2 \bar\theta^2 \p^2 \phi
\ee
and satisfies 
\be
L=\bar L\, , ~~~~D^2L=0 .
\ee
This implies that  the vector $H_\mu$ is divergeneless  
\be
\p^\mu H_\mu=0.
\ee
The action (\ref{m3}) can be written as
\beq
L=\int d^4 \theta \left( -L^2 +  {1\over 64 \Lambda^4} X\bar X \right)= 
\int d^4 \theta \left({1\over 64 \Lambda^4}X\bar X  \right) + \left(  {1\over 4} \int d^2\theta X +h.c. \right),
\eeq{m5}
with 
\be
\bar{X}\equiv D^\alpha L D_\alpha L= \frac{1}{2}D^2 L^2.
\ee
Note that $\bar{X}$ is antichiral, so $X$ is chiral and obeys 
$X^2=0$. Then the  Lagrangian~(\ref{m5}) is the same as in~\cite{ks,feruglio}  (modulo normalization factors). 
In particular, $X$ is explicitly written in chiral coordinates as 
\beq
X=\bar{D}_{\dot \alpha}L\bar{D}^{\dot \alpha}L= \bar\psi ^2 - 2 \theta \sigma_\mu \bar \psi ( i \p^\mu \phi + H^\mu) 
+ \theta^2 [ 2 i \p^\mu \psi \sigma_\mu \bar\psi +  (  i \p^\mu \phi + H^\mu )^2 ]
\eeq{sm4}
 therefore, it is chiral with 
 auxiliary field ${\cal F}$
\be{\cal F}= (i \p_\mu \phi + H_\mu) (i \p^\mu \phi + H^\mu). 
\ee
The goldstino now is given by
\be
G_{\alpha}= - 2 \sigma_{\mu \alpha \dot{\alpha}} \bar \psi ^{\dot{\alpha}} ( i \p^\mu \phi + H^\mu).
\ee
It is easy to see that the bosonic part of (\ref{m5}) is
\be
{\cal L}^B=\frac{1}{2}H_\mu H^\mu-\frac{1}{2}\partial_\mu \phi \partial^\mu \phi 
+\frac{1}{64 \Lambda^4} |( i \partial_\mu \phi +H_\mu)^2|^2
\ee
There is a supersymmetric vacuum $H_\mu=0$, $\phi=const.$ and a supersymmetry breaking one (with $\phi=const.$ 
) 
\be
H_\mu H^\mu= -16 \Lambda^4. \label{p}
\ee
In this case, supersymmetry is broken and the theory reduces to the standard
Volkov-Akulov for the goldstino $G$.  
In spite of appearances, the vacuum solution (\ref{p}) does not breaks Lorentz invariance, 
since the divergenceless vector $H_\mu$ and $\partial_\mu \phi$ combine into the unconstrained vector $A_\mu$,
which does not propagate, because it has algebraic equations of motion. Therefore, a nonzero constant vev for $A_\mu$ 
does not affect the dynamics since it either disappears from the Lagrangian or it arranges itself into Lorentz-invariant 
composite quantitites.
We also note that, after using (\ref{sm4}), the action (\ref{m5}) 
is written entirely in terms of the goldstino field
$G_\alpha$. Again here, similarly to  the spinor superfield case above, 
there is no superpartner of the goldstino. The sgoldstino is decoupled as all fields 
before susy breaking were auxiliaries and therefore (\ref{m5}) may be consider as the zero-momentum limit of a theory were 
these were propagating. In this limit, the sgoldstino decouples and the theory describes a Volkov-Akulov model.

\section{Validity of the Volkov-Akulov Description}

The theories above, as well as the one we will examine later, 
must be understood as effective IR theories. If a supersymmetric UV completion existed, 
then the sgoldstino $\varphi$ would have a large but finite mass $m_s$. 
It would interact with the goldstino through terms of the schematic form 
\beq
\kappa G_\alpha G^\alpha \varphi + (m_s^2/2)\varphi^2+...,
\eeq{sgd1}
 with a coupling constant $\kappa=O(m_s^2/f)$. 
 At energies below $m_s$, the sgoldstino fields can be integrated out, producing additional irrelevant 
 operators weighted by inverse powers of the new scale $\Lambda'=f/m_s$. 
 Curiously, these additional interactions become negligible when the sgoldstino is 
 massive but {\em lighter} than $\sqrt{f}$: $\Lambda' \gg \sqrt{f} \rightarrow m_s \ll \sqrt{f}$. 
 We will explicitly demonstrate this in the case of supersymmetric  theories with  chiral multiplets.

Let us recall that in globally supersymmetry theory 
with $n+1$ chiral multiplets $\Phi^i$, the Yukawa couplings arise from the term
\be
{\cal L}\supset  W_{ij}(\phi)\chi^i\chi^j+h.c\, , ~~~~i,j=0,1,...,n ,
\label{yuk}
\ee
where $\phi^i,\chi^i$ are the scalars and fermions of the chirals and $W_{ij}=\partial^2 W/\partial\phi^i\partial\phi^j$.
The potential is 
\be
V=W_i W^i ,
\ee
where the notation $W^i=(W_i)^\dagger$ is used
and let us assume for the moment that the K\"ahler metric is flat. 
The values of the fields in the ground state are $\big<
\phi^i\big>=a^i,~\big<F^i\big>=f^i,~ \big<\psi_i\big>=0$ and the equation of motions give 
\be
\bar{f}_i=-w_i\, , ~~~~w_{ij}f^j=0,
\ee
where 
\be
w_i=W_i(a^i)\, , ~~~w_{ij}=W_{ij}(a^i)\, , ~~~~...
\ee
The term (\ref{yuk}) gives then rise to the interaction
\be
{\cal L}\supset  w_{ijk}\delta\phi^k\chi^i\chi^j+h.c , \label{yuk1}
\ee
where $\delta\phi^i=\phi^i-a^i$.
Since supersymmetry is broken, the fermionic shifts will not vanish in the vacuum 
\be
\label{fi}
<\delta \chi_i> = -f_i \epsilon.
\ee
By an appropriate rotation  of $\chi_i$, we can
define new fermionic fields $\tilde{\chi}_i$ 
\be
\tilde{\chi}_i={R_i}^j\chi_j ,
\ee
where ${R_i}^j$ is an appropriate matrix such that the non-zero fermionic shift  are along a specific direction,
which we will call it (``0'')
\be
<\delta \tilde{\chi}_{0}> = -f \epsilon\, , ~~~~<\delta \tilde{\chi}_{a}> = 0  \, , ~~~a=1,\ldots,n ,
\ee
with $|f|^2=f_if^i$. Clearly $\tilde{\chi}_0$ is the goldstino, which is defined then as
\be
\tilde{\chi}_0={R_0}^i \delta \chi_i\,
\ee
and the rest of the modes are given by
\be
\delta \tilde{\chi}_a={R_a}^i \delta \chi_i .
\ee
The matrix $R_{ij}$ is orthogonal  and chosen to satisfy
\be
{R_a}^i f_i=0 \, .\label{R1}
\ee
When this equation is satisfied, then ${R_0}^i=f_i/|f|$ so that the goldstino is 
\be
\delta \tilde{\chi}_0=\frac{f_i}{|f|} \delta \chi_i\, .\label{xrot}
\ee
Note that instead of rotating $\chi_i$'s, we could have rotated the original superfields $\Phi^i$ 
so that the goldstino belongs  to the  $\tilde{\Phi}^0$ goldstino superfield,
which is a linear combination of the original fields. According to (\ref{xrot}),  $\tilde{\Phi}^0$ is 
\be
 \tilde{\Phi}_0=\frac{f_i}{|f|} \Phi^i .
 \ee
 The rest of the superfields are given by
 \be
 \tilde{\Phi}_a={R_a}^i\Phi^i  \label{rf};
 \ee
therefore, the sgoldstino is 
\be
\phi^0=\frac{f_i}{|f|} \phi^i .
\ee
 
 The interaction (\ref{yuk1}) is written then in terms of the 
 new fields as 
 \be
{\cal L}\supset {R^i}_n{R^j}_m{R^k}_lw_{ijk}\delta\tilde{\phi}^n\tilde{\chi}^m\tilde{\chi}^l .
\ee
The possible Yukawa coupling of the golstino  are
\be
&&{\cal L}_1\supset {R^i}_0{R^j}_0{R^k}_0w_{ijk}\delta\tilde{\phi}^0\tilde{\chi}^0\tilde{\chi}^0=|f|^{-3}
f^if^jf^kw_{ijk}\delta\tilde{\phi}^0\tilde{\chi}^0\tilde{\chi}^0=
|f|^{-3}
\, s\, \delta\tilde{\phi}^0\tilde{\chi}^0\tilde{\chi}^0\\
&&{\cal L}_2\supset {R^i}_a{R^j}_0{R^k}_0w_{ijk}\delta\tilde{\phi}^a\tilde{\chi}^0\tilde{\chi}^0=
|f|^{-2}{R^i}_af^jf^kw_{ijk}\delta\tilde{\phi}^a\tilde{\chi}^0\tilde{\chi}^0=
|f|^{-2}{R^i}_a\, s_i\, \delta\tilde{\phi}^a\tilde{\chi}^0\tilde{\chi}^0\\
&&{\cal L}_2\supset {R^i}_a{R^j}_b{R^k}_0w_{ijk}\delta\tilde{\phi}^a\tilde{\chi}^b\tilde{\chi}^0 ,
\ee
where 
\be
s=f^if^jf^kw_{ijk}\, , ~~~~s_k=f^if^jw_{ijk} .
\ee
We will show now that 
\be
s=0\, , ~~~~s_i=0
\ee
so that  a  globally supersymmetric theory 
the only trilinear Yukawa coupling is the one that contains only one goldstino or one sgoldstino. 
For this, we need to recall
that the fermionic mass matrix $m_F=w_{ij}$ has a zero eigenvalue 
\be
{m_F}_{ij}f^j=0\, ,   \label{mf}
\ee
and the  bosonic mass matrix 
\be 
M_B^2=\left(\begin{array}{cc}
                                    m_F^\dagger m_F&  \sigma\\
                                    \sigma^\dagger&m_Fm_F^\dagger
                                   \end{array}\right)\, \, , ~~~~~\sigma_{ij}=w_{ijk}f^k
                                   \ee
is positive definite
\be
\big<\Psi |M^2_B|\Psi\big>\geq 0  . \label{pm}
\ee
For 
\be
|\Psi\big>=\left(\begin{array}{c}
                  f_i\\
                  f^i
                 \end{array}
\right)
\ee
we get, since $m_F$ annihilates $f^i$,
\be
\operatorname{Re}(f^if^js_{ij})\geq 0 .
\ee
Moreover, since  $m_F$ annihilates also $e^{i\varphi}f^i$, where $\phi$ is an arbitrary phase, we get in 
general
\be
\operatorname{Re}(e^{2i\varphi}f^if^j\sigma_{ij})\geq 0
\ee
which leads to 
 \be
s= f^if^j\sigma_{ij}=f^if^jf^kw_{ijk}=0 . \label{s0}
 \ee
 Therefore, the coupling ${\cal L}_1$ vanishes and  there is no (goldstino$^2$\, sgoldstino)  coupling. 

 We can also prove that there is no (goldstino$^2$\, scalar) Yukawa coupling
 by showing that $s_i=0$, which means that ${\cal L}_2$ vanishes as well.  
 By using (\ref{s0}), it is easy to see that in fact
 \be
\big<\Psi |M^2_B|\Psi\big>= 0
\ee
and since $M_B^2$ is positive definite, $M_B^2$ annihilates $|\Psi\big>$
\be
M_b^2|\Psi\big>=0 .
\ee
Then, by using (\ref{mf},\ref{s0}), we find  
\be
\sigma_{ij}f^j=w_{ijk}f^jf^k=0 .
\ee 
Therefore, $s_{i}=0$ and the interaction ${\cal L}_2$ similarly vanish. As a result, in a 
 globally supersymmetric theory, the only  Yukawa coupling that is allowed, 
is only ${\cal L}_3$, i.e., a single goldstino interacting with a scalar and a fermion 
of the matter scalar multiplet or a single sgoldstino interacting with two fermions of the matter scalar multiplet.  
In particular, this means that there is no way to break supersymmetry just with 
a single chiral multiplet. 

\vskip.2in 

Let us now turn to the general case of a non-flat K\"ahler metric $g_{i\bar j}$.  
In this case, the bosonic mass matrix is 
\be
\label{BM}
M^{2}_{B}=
\left( \begin{array}{ll}
-{K^j}_i+{(m_F^\dagger m_F)^j}_i 
& 
\  \sigma
\\ 
\ \ \ \ \ \ \ \ \ \ \ \sigma^\dagger
& 
-{K_i}^j+{(m_F^\dagger m_F)_i}^j 
\end{array}\right)\, .
\ee
where 
\be
{K^j}_i=K_{\bar{j}i}=K_{\bar{j}i\bar{m}n}\bar{f}^{\bar{m}}f^k
\ee
and $K_{\bar{j}i\bar{m}n}=R_{\bar{j}i\bar{m}n}$ in normal coordinates.
Now, the corresponding  relation (\ref{pm}) for the positivity of $M_B^2$ does not lead to any conclusive relation. 
The Yukawa couplings originate from the term
\be
{\cal L}\supset \Big(W_{ij}-\Gamma_{ij}^k W_k\Big)\chi^i\chi^j +h.c.
\ee
 which  gives rise to 
 \be
 {\cal L}\supset \Big(W_{ijk}-\partial_k\Gamma_{ij}^l W_l-\Gamma_{ij}^lW_{lk}\Big)\delta\phi^k\chi^i\chi^j +h.c..
\ee
Rotating the fields such that again the goldstino  is in the 0-direction as before, we get the 
interaction 
\be
 {\cal L}\supset \tilde{s}\, \delta\phi^0\chi^0\chi^0+h.c.
\ee
where 
\be
\tilde s=(W_{ijk}-\partial_k\Gamma_{ij}^l W_l-\Gamma_{ij}^lW_{lk})f^if^jf^k .
\ee
Clearly now $\tilde s\neq 0$ as can easily be checked  for the simplest case of a linear superpotential $W=f \Phi$. 
In fact 
it is easy to see that if the scale of the K\"ahler manifold is $\Lambda$ 
then the sgoldstino mass is 
\be
m_{s}\sim \frac{f}{\Lambda}
\ee 
and $\tilde s$ is of the order of 
\be
\tilde s\sim \frac{f}{\Lambda^2}\sim \frac{m_s^2}{f} .
\ee
Therefore, the effective coupling in the IR will be schematically of the form 
\be
\frac{m_s^2}{f} \chi^0\chi^0\phi^0-\frac{1}{2}m_s^2\phi_0^2+\cdots +h.c
\ee
which gives rise to a term of the form 
\be
{\cal L}\supset \frac{m_s^2}{f^2}(\chi^0\bar{\chi}^0)^2 \label{yk}
\ee
after integrating out the sgoldstino. Such a term is supressed by the scale $\Lambda'=f/m_s$ and therefore
it can be ignored as long as it is much larger than the Volkov-Akulov scale $\sqrt{f}$  ($\Lambda'>> \sqrt{f}$). 
In this 
case, interactions like  (\ref{yk}) can safely be ignored and the theory 
will be described by  
Volkov-Akulov for 
\be
\frac{f}{m_s}>>\sqrt{f} .
\ee
In other words, the Volkov-Akulov description is valid for 
\be
m_s<<\sqrt{f}<<\Lambda .
\ee
This limit is the one considered in the models with constraint superfields in which the sgoldstino can be safely integrated out 
resulting in a non-linearly realized supersymmetric Volkov-Akulov theory for the goldstino mode. 
The V-A description is then valid only up to a UV
cutoff equal to the mass $m_{lightest}$  of the lightest particle mixing with the
goldstino. This particle can be the sgoldstino or one of the fermions
orthogonal to the goldstino.
Of course, as in all effective Lagrangians, the V-A scale $f$ must obey $f > m_{lightest}^2$.

\section{The Complex Linear Multiplet}
We have explicitly demonstrated in the previous section that higher dimensional operators contribute 
to the vacuum structure and may lead to supersymmetry breaking.

Here we will see that it is possible to break supersymmetry without intorducing any Lorentz non-invariant vev.

%
The reason that the potential~(\ref{vf}) cannot break superymmetry  is that the two terms in (\ref{vf}), coming from the two- and four- derivative
terms of (\ref{f11}) have the same sign. Clearly, new extrema can emerge only if these terms have opposite sign,
i.e. if the first contribution coming form the leading term in (\ref{f11}) flips sign. This can happen for the 
complex linear multiplet \cite{gates1,gates2}. 

The complex linear or nonminimal multiplet is defined as 
\be
\label{L1}
\bar{D}^2\S=0.
\ee
The constraint (\ref{L1}) above is just the field equation for a free chiral multiplet.  
Note that if the further constraint $\S=\bar \S$ is imposed, the complex linear multiplet turns into a linear one.  
The standard kinetic Lagrangian for the complex linear superfield in superspace reads
\be
\label{Lang1}
{\cal L}_0=-\int d^4\theta\,  \S\bar{\S}.
\ee
Note the relative minus sign compared to the kinetic Lagrangian of a chiral multiplet. This is necessary for the theory to contain no ghosts. 
The relative minus sign of the complex linear multiplet $\S$ 
compared to the standard kinetic term for a chiral multiplet $\Phi$ can be understood 
in terms of  a duality transformation. Indeed, consider 
 the action
\be
\label{Lang11}
{\cal L}_D=-\int d^4\theta \big{(}\S\bar{\S}+\Phi \Sigma+\bar{\Phi}\bar{\Sigma}\big{)},
\ee
where $\Phi$ is chiral and $\S$ is unconstrained. Integrating out $\Phi$ we get both eq.~(\ref{Lang1}) and the constraint 
(\ref{L1}). However, by integrating out $\Sigma$, we get $\S=-\bar \Phi$. Plugging back this equality into (\ref{Lang11}), we get the 
standard kinetic term of a chiral multiplet
\be
\label{Lang110}
{\cal L}_0=\int d^4\theta \Phi\bar{\Phi}.
\ee
As announced, the overall sign in Lagrangian~(\ref{Lang110}) is opposite to that of (\ref{Lang1}).

To find the superspace equation of motion, we should express $\Sigma$ in terms of an unconstrained superfield.
This can be done by introducing a general spinor superfield $\Psi^\alpha$ with gauge transformation
\be
\delta \Psi_\alpha=D^{\beta}\Lambda_{(\alpha\beta)}
\ee
where $\Lambda_{(\alpha\beta)}$ is arbitrary. It is easy to see that by defining 
\be
\S=\bar D _{\dot \alpha}\bar \Psi ^{\dot \alpha} \label{psi},
\ee
$\S$  satisfies the constraint (\ref{L1}).  Then the field equation
following from  eq.~(\ref{Lang1}) is 
\be
D_\alpha\Sigma=0\, .
\ee
Therefore, the field equation of a complex linear multiplet is just the constraint of a chiral multiplet and, as noticed above, the 
constraint on a linear is the field equation of a chiral. This indicated the duality between the two kind of 
multiplets, at least in the free case.  The field content of the complex linear multiplet $\S$
is revealed  via the projection over components as
\begin{align}
\nn 
A&=\S |,
\\
\nn
\psi_{\alpha}&=\frac{1}{\sqrt{2}}D_{\alpha}\bar{\S}|,
\\
\nn
F&=-\frac{1}{4}D^2\S|,
\\
\nn
\lambda_{\alpha}&=\frac{1}{\sqrt{2}}D_{\alpha}\S|,
\\
\nn
P_{\alpha\dot{\beta}}&=\bar{D}_{\dot{\beta}}D_{\alpha}\S|\, , ~~~~~~~\qquad
\bar P_{\alpha\dot{\beta}}=-D_\beta \bar{D}_{\dot{\alpha}}\bar \S|,
\\
\label{comp}
\chi_{\alpha}&=\frac{1}{2}\bar D_{\dot{\alpha}}D_{\alpha}\bar D^{\dot{\alpha}}\bar \S|\, , 
~~~~~~~ \bar{\chi}_{\dot{\alpha}}=\frac{1}{2}D^{\alpha}\bar{D}_{\dot{\alpha}}D_{\alpha}\S|.
\end{align}
In other words, a complex linear multiplet contains a chiral multiplet $(A,\lambda_\alpha,F)$ and an antichiral 
spinor superfield ($\psi_\alpha,P_{\alpha\dot{\beta}},\chi_\alpha)$. 
Therefore, the complex linear multiplet is  a reducible 
12 + 12 dimensional representation of the  ${\cal N} =$ 1 supersymmetry. 
It should be noted that since $\S$ is not  chiral, there is no superpotential and there are no supersymmetric non-derivative interactions. 
However, the complex linear multiplet can still be consistently coupled to ordinary 
vector multiplets of the ${\cal N}=1$ theory.

We give for later use the supersymmetry transformations of the fermionic components of $\Sigma$
\begin{align}
\label{susy1}
\delta\psi_{\alpha}&=\sqrt{2} i \s^{\mu}_{\alpha\dot{\beta}}\bar{\xi}^{\dot{\beta}}\p_{\mu}\bar{A}
-\frac{1}{\sqrt{2}}\bar{\xi}^{\dot{\beta}}\bar{P}_{\alpha\dot{\beta}}
\\
\delta\chi_{\alpha}&=
2 i \s^{\nu}_{\alpha \dot{\alpha}} \bar{\s}^{\mu \dot{\alpha} \beta} \xi_\beta
\p_\mu \bar{P}_\nu
+i \s^{\mu}_{\alpha \dot{\alpha}} \bar{\s}^{\nu \dot{\alpha} \beta} \xi_\beta
\p_\mu \bar{P}_\nu
-4 \xi_{\alpha} \p^2 \bar{A} + 2 i \s^{\mu}_{\alpha \dot{\alpha}}
\bar{\xi}^{\dot{\alpha}} \p_{\mu} \bar{F}  \label{susy2}\\
\delta\lambda_{\alpha}&= \sqrt{2}\xi_{\alpha}F-\frac{1}{\sqrt{2}}\bar{\xi}^{\dot{\beta}}P_{\alpha\dot{\beta}}. \label{susy3}
\end{align}
The transformation rules of the bosonic sector of the complex linear multiplet
are
\be
\label{DB1}
\delta A &=& \sqrt{2} \bar \xi \bar \psi + \sqrt{2}  \xi  \lambda ,
\\
\label{DB2}
\delta F&=& \frac{i}{\sqrt{2}} \bar \xi \bar \s ^\mu \p_\mu \l 
+\frac{1}{2} \bar \xi \bar \chi ,
\\
\label{DB3}
\delta P_{\alpha \dot{\beta}} &=&
- 2 \sqrt{2} i \xi^{\gamma} \s^{\mu}_{\gamma \dot{\beta} } \p_\mu \l_{\alpha}
+  \sqrt{2} i \xi_{\alpha} \s^{\mu}_{\beta \dot{\beta} } \p_\mu \l_{\beta}
-\xi_{\alpha} \bar \l_{\dot{\beta}}
-2 \sqrt{2} i \bar \xi_{\dot{\beta}} \s^{\mu}_{\alpha \dot{\rho} }  \p_\mu \bar \psi^{\dot{\rho}}.
\ee
In terms of the components of $\S$,  Lagrangian~(\ref{Lang1}) is explicitly written as  
\be
\label{Lang1comp1}
{\cal L}_0=A\p^2\bar{A}-F\bar{F}
+i  \p_\mu \bar \psi \bar \sigma ^\mu \psi
+\frac{1}{2}P_{\mu}\bar{P}^{\mu}+\frac{1}{2\sqrt{2}}\big{(}\chi\lambda+\bar{\chi}\bar \lambda\big{)}.
\ee
The complex vector $P_{\mu}$, the complex scalar $F$ and the spinors $\lambda,~\chi$ are auxiliary fields.
Note that the minus sign in front of the superspace action (\ref{Lang1}) guarantees that the scalar $A$ is a normal field
and not a ghost. However, this choice of sign has flipped the sign of the $F\bar F$ relative to the action for 
a chiral multiplet. This flip of sign is  of fundamental importance for what follows and  
leads to supersymmetry breaking.

\section{SUSY Breaking by Complex Linear Multiplets}

As we have noticed before, although one can couple the linear multiplet to gauge fields 
\cite{deo,penati,proyen,grassi,tarta}, one cannot write down mass terms or 
non-derivative interactions as in the chiral multiplet case by means of a superpotential. 
So, the best we can hope for is to introduce a potential indirectly by using the higher dimensional 
operators first discussed in \cite{fer}.  The idea of \cite{fer} has been recently revisited and the  
emergent potential   for chiral and vector multiplets has been discussed in~\cite{ovrut,Far1,Far2}.

To achieve this, we 
introduce the following Lagrangian in superspace
\be
\label{ep1}
{\cal L}_{EP}=\int d^4\theta \  \frac{1}{64\, \Lambda^4}\, \   
D^{\alpha}\S D_{\alpha}\S \bar{D}_{\dot{\alpha}}\bar{\S}\bar{D}^{\dot{\alpha}}\bar{\S},
\ee
where $\Lambda$ is a mass scale. Then, the theory is described by 
\begin{align}
{\cal{L}}_\S&={\cal L}_0+{\cal L}_{EP}\nonumber \\
&=\int d^4\theta \ \left(-\S\bar{\S}+ \frac{1}{64\Lambda^4} \   
D^{\alpha}\S D_{\alpha}\S \bar{D}_{\dot{\alpha}}\bar{\S}\bar{D}^{\dot{\alpha}}\bar{\S}\right).  \label{Lo}
\end{align}
By using the unconstrained superfield $\Phi_\alpha$, we find that the field equations are 
\be
D_\alpha \S+\frac{1}{32\Lambda^4}D_{\alpha}\bar{D}_{\dot{\alpha}}\left(D^{\beta}\S D_{\beta}\S \bar{D}^{\dot{\alpha}}\bar{\S}
\right)=0. \label{sol}
\ee
Clearly, the above equation always admits the supersummetry preserving solution
\be
D_\alpha \S=0 \, .
\ee
We are interested to investigate if another, supersymmetry breaking solution to (\ref{sol}) exists. 

The component form of the bosonic part of  eq.~(\ref{ep1}) is 
\be
\label{ep2}
{\cal L}^B_{EP}=\frac{1}{64\Lambda^4}\Big{(} P^{\mu} P_{\mu} \bar{P}^{\nu} \bar{P}_{\nu} 
+4P_{\mu}\bar{P}^{\mu}F\bar{F}+16   F^2\bar{F}^2\Big{)},
\ee
so that  the bosonic part of the full  Lagrangian (\ref{Lo}) turns out to be
\begin{align}
\nn
{\cal L}^B=
&-F\bar{F}+A\p^2\bar{A}+\frac{1}{2}P_{\mu}\bar{P}^{\mu}
\\
\label{Lang2}
&+\frac{1}{64\Lambda^4}\Big{(} P^{\mu} P_{\mu} \bar{P}^{\nu} \bar{P}_{\nu} 
+4P_{\mu}\bar{P}^{\mu}F\bar{F}+16   F^2\bar{F}^2\Big{)}.
\end{align}
From the equations of motion for the complex auxiliary vector we find that 
\be
\label{P=0}
P_{\mu}=0,
\ee
whereas the equations of motion for the auxiliary scalar turns out to be 
\be
\label{FF}
F\left(1-\frac{1}{2\Lambda^4}F\bar{F}\right)=0.
\ee
There are now two solutions:
\be
&&(i) ~~~~F=0\, ,\label{v}\\
&&(ii)~~F\bar F
=2\Lambda^4 . \label{v2}
\ee
Clearly, as it follows from eqs.~(\ref{susy1},\ref{susy2},\ref{susy3}), the first vacuum $F=0$ is the supersymmetric one,
where supersymmetry is exact. However, the second vacuum, described by 
the solution (\ref{v2}), explicitly breaks supersymmetry. 
We note that the theories with $F=0$ and $F\neq 0$
should not be thought as phases of the same theory but rather as two different theories. 
This can be illustrated by the following example.
Consider a scalar A and an auxiliary field $Y$ with Lagrangian:
\be
{\cal L}_{AY}=-\frac{1}{2}\partial_\mu A \partial^\mu A -
\frac{1}{2}Y^2(a A^2 +b) +\frac{1}{4}Y^4.
\ee
Solving for $Y$ we get two solutions: $Y=0$, which gives the free scalar Lagrangian 
\be
{\cal L}_{A}=- \frac{1}{2} \partial_\mu A \partial^\mu A, 
\ee
and 
\be
Y^2=aA^2 +b,
\ee
which gives the interacting Lagrangian 
\be
{\cal L}'_A= -\frac{1}{2}\partial_\mu A \partial^\mu A -\frac{1}{4}(aA^2 +b)^2.
\ee
No transition either perturbative or nonperturbative can occur between the two, 
precisely because the equations for $Y$ are algebraic, so they are truly two different theories.

It should also be noted that the susy-breaking vacuum is specified by the modulus of the auxiliary field $F$. So, $F$ itself 
is specified only up to a phase. This is expected due to the invariance 
of Lagrangian~(\ref{Lo}) under the global $U(1)$ transformation
\be
\S\to e^{i\phi}\S.
\ee

For completeness, we give the component form of Lagrangian~(\ref{Lo})
\allowdisplaybreaks
\begin{align}
\label{Lang1comp}
{\cal L}_\S&=A\p^2\bar{A}-F\bar{F}
+i  \p_\mu \bar \psi \bar \sigma ^\mu \psi
+\frac{1}{2}P_{\mu}\bar{P}^{\mu}+\frac{1}{2\sqrt{2}}\big{(}\chi\lambda+\bar{\chi}\bar \lambda\big{)}
\nn \\
& + \frac{1}{64\Lambda^4}\Big{\{}  4 (\l^\alpha \p^2 \l_\alpha ) \bar{\l}^2 
+ 2 \sqrt{2} i (\p_\mu \bar{\chi} \bar{\s}^\mu \l ) \bar{\l}^2  
\\
\nn
& - 16 F \p^2 A \bar{\l}^2 
+ 8 i F \p^\mu P_\mu  \bar{\l}^2
\\
\nn
& +8 \p^2 A \bar{\l} \bar{\s}^\kappa \l \bar{P}_\kappa 
+4i \bar{\l} \bar{\s}^\kappa \s^\nu \bar{\s}^\mu \l \bar{P}_\kappa \p_\mu P_\nu
\\
\nn
&+8i \bar{\l} \bar{\s}^\kappa \s^\mu \p_\mu \bar{\psi} F \bar{P}_\kappa
-16 \p_\mu \bar{\psi} \bar{\s}^\mu \l \p_\nu \psi \s^\nu \bar{\l} 
+ 4 i \p_\mu \bar{\psi}  \bar{\s}^\mu \l \bar{P}^2
\\
\nn
&+\frac{1}{2} \Omega^{\beta \dot{\beta} \alpha} \Omega_{\beta \dot{\beta} \alpha} \bar{\l}^2
-8i \bar{\l}^2 P^\kappa \p_\kappa F 
\\
\nn
&+ \sqrt{2}\bar{P}_\mu \bar{\l}_{\dot{\alpha}} 
\bar{\s}^{\mu \dot{\alpha} \beta} \Omega_{\beta \dot{\beta} \alpha} 
\bar{\s}^{\kappa \dot{\beta} \alpha} P_\kappa 
+ 4 i P^2 \p_\mu \psi \s^\mu  \bar{\l} + P^2 \bar{P}^2
\\
\nn
&- 8 \sqrt{2} F \bar{\chi} \bar{\l} \bar{F} 
- 8 F \bar{F} P_\nu \bar{P}^\nu 
- 2 \sqrt{2} \chi \s^\mu \bar{\l} P_\mu F
\\
\nn
&+4i F P_{\mu} \bar{\l}  \bar{\s}^{\mu} \s^\nu \p_\nu \bar{\l}
-16 i \l \s^\nu \bar{\l} \bar{F} \p_\nu F
\\
\nn
&+ 2 \sqrt{2}\bar{P}_\nu \bar{\s}^{\nu \dot{\beta} \beta} \Omega_{\beta \dot{\beta} \alpha} \l^\alpha  \bar{F}
-2  \Omega_{\beta \dot{\beta} \alpha} \chi^\beta \bar{\l}^{\dot{\beta}}  \l^\alpha
+2 \sqrt{2} i \p_\mu \bar{\l}_{\dot{\rho}} \bar{\s}^{\mu \dot{\rho} \beta}  
\Omega_{\beta \dot{\beta} \alpha} \l^\alpha \bar{\l}^{\dot{\beta}} 
\\
\nn
&- 8 i \p_\nu \psi \s^\nu \bar{\s}^{\mu} \l P_\mu \bar{F}
- \sqrt{2} \l \s^\mu \bar{\s}^{\nu} \chi P_{\mu} \bar{P}_{\nu}
- 2 i \l \s^\kappa \bar{\s}^{\mu}  \s^\nu \p_\nu \bar{\l} P_{\kappa} \bar{P}_{\mu}
\\
\nn
&- 8 \l \s^\nu \bar{\l} P_\nu \p^2 \bar{A} 
- 8 i  \l \s^\nu \bar{\l} P_\nu \p_\mu \bar{P}^\mu
\\
\nn
&+ 16 F^2 \bar{F}^2 
-8 \sqrt{2} \l \chi F \bar{F} 
- 16 i \l \s^\nu \p_\nu \bar{\l} F \bar{F} 
\\
\nn
&- 16 \l^2 \bar{F}\p^2 \bar{A} 
- 16 i \l^2 \bar{F} \p_\mu \bar{P}^\mu
- \l^2 \Xi^2\Big{\}}  ,
\end{align}
where 
\be
\Omega^{\beta \dot{\beta} \alpha} =
-2 \sqrt{2} i \bar{\s}^{\mu \dot{\beta} \beta}\p_\mu \l^\alpha
- i \sqrt{2} \epsilon^{\beta \alpha }  \bar{\s}^{\mu \dot{\beta} \gamma}
\p_\mu \l_\gamma
- \epsilon^{\beta \alpha } \bar{\chi}^{\dot{\beta}} \ \ , \ \ 
\Omega_{\rho \dot{\rho} \s} 
= \epsilon_{\rho \beta} \epsilon_{\dot{\rho} \dot{\beta}} \epsilon_{\s \alpha} 
\Omega^{\beta \dot{\beta} \alpha} 
\ee
and
\be
\Xi_{\beta} = \chi_{\beta} + \sqrt{2} i \s^\nu_{\beta \dot{\beta}} \p_\nu
\bar{\l}^{\dot{\beta}}.
\ee

We should note that Lagrangian~ (\ref{Lang1comp}) contains also  first derivatives of the auxiliaries $F,P_\mu,
\chi$. Therefore, one
may question if these fields are really auxiliaries. However, it can easily be checked that these derivative terms
are always multiplied by fermions. Therefore their equations of motion can be integrated by iteration in a 
power series of the 
fermions, which terminates due to the nilpotent nature of the latter.

To identify the goldstino mode, one should look at the supersymmetry transformations and, in particular, to the 
fermion shifts. It is then easy to recognize that since 
\be
\delta \lambda_\alpha=2 \, \xi_\alpha \Lambda^2+\ldots \, ,
\ee
the goldstino of the broken supersymmetry is proportional to  $\lambda$, i.e.,   one of the auxiliary fermions.
Here something unusual has happened; namely,
an auxiliary fermion has turned into a goldstino mode in the susy breaking vacuum. 
However, the latter is  propagating and 
the vacuum (\ref{v2})  should definitely give rise to a kinetic term for $\lambda$.  
Indeed, it is straightforward to see that the higher dimensional operator Lagrangian
gives rise to the following coupling for the auxiliary fermion $\lambda$
\be
\label{goldstino}
{\cal L}_{EP} \supset (\frac{1}{4\Lambda^4} F \bar{F})\ i \p_{\mu} \bar{\lambda}
\bar{\s}^{\mu} \lambda.
\ee
In the susy breaking vacuum obtained from eq.~(\ref{FF}) we have
\be
\label{vac}
< F \bar{F}> = 2\Lambda^4,
\ee
leading to a standard fermionic kinetic term with the correct sign
\be
\label{goldstino-vac}
{\cal L}_{EP} \supset \frac{i}{2} \p_{\mu} \bar{\lambda} \bar{\s}^{\mu} \lambda.
\ee
Therefore, on the susy breaking vacuum (\ref{v2}), the auxiliary fermion 
$\lambda$ is propagating and it is proportional to the goldstino mode of broken susy. 
Note that due to the model independent  relation (\ref{vac}),
the kinetic term (\ref{goldstino-vac}) for the goldstino is also model
independent. 
In fact what has happened here is that the susy breaking phase
is a realization of non-linear supersymmetry.

We  should also mention that the fermion bilinears $\chi\lambda$ and $\bar \chi \bar \lambda$ appear in the action 
as
\be
{\cal L}_\Sigma \supset \frac{1}{2\sqrt{2}}\left(1-\frac{F\bar F}{2\Lambda^4}\right)\Big{(}
\chi\lambda +\bar \chi \bar \lambda\Big{)}.
\ee
Such terms vanish on the non-supersymmetric vacuum and protect the theory from unwanted, dangerous terms.
Moreover, as in the spinor superfield and real multiplet case,  there is no superpartner of the goldstino. 
In fact, the propagating modes are the real scalar $A$, the fermion $\psi$ and the golstino $\lambda $, which 
definitely do not form a multiplet of the (broken) susy. The reason again is that  the rest of the fields 
of the complex linear multiplet are auxiliaries
and therefore the sgoldstino decouples.

One could proceed and solve the field equations for the auxiliaries in (\ref{Lang1comp}). 
Although this is a formidable task,   
there is an indirect way to proceed in superspace. We will show below that the theory  
(\ref{Lang1comp}) describes a free chiral 
multiplet and a constraint chiral superfield which describes a Volkov-Akulov mode.  
To see how this happens, 
let us remind briefly some aspects of non-linear supersymmetry realizations.
It is well known that the following Lagrangian~\cite{feruglio}
\be
\label{off-shell}
{\cal L}=\int d^4 \theta \, X_{NL} \bar X_{NL}
+ \sqrt{2} \Lambda^2 \left(\int d^2\theta\, X_{NL}+h.c\right) 
+ \left(\int d^2\theta\, \Psi X_{NL}^2 +h.c\right)
\ee 
is on-shell equivalent to the Akulov-Volkov theory.  
In fact, the Lagrange multiplier chiral superfield $\Psi$  imposes the constraint
\be
X_{NL}^2=0
\ee
on the chiral superfield $X_{NL}$, leads to the non-linear realization of supersymmetry 
\cite{rocek,feruglio,ks} and  reproduces the Volkov-Akulov model.
The Lagrangian (\ref{off-shell}) gives rise to the following two equations of motion in superspace
\be
\label{DDX}
-\frac{1}{4} \bar D ^2 \bar X_{NL} +  \sqrt{2} \Lambda^2  + 2\Psi X_{NL} &=&0,
\\
\label{XX}
X_{NL}^2=0. \ \ \ \ \ \ \ \  &&
\ee
The theory we consider here is described by the Lagrangian 
\be
{\cal L}= -\int d^4\theta\,  \S\bar{\S} + \int d^4\theta \  \frac{1}{64\, \Lambda^4}\, \   
D^{\alpha}\S D_{\alpha}\S \bar{D}_{\dot{\alpha}}\bar{\S}\bar{D}^{\dot{\alpha}}\bar{\S}
\ee
and the  superfield equations of motion are written as
\be
D_\alpha \S +\frac{1}{32\Lambda^4}D_{\alpha}\bar{D}_{\dot{\alpha}}\left(D^{\beta}\S D_{\beta}\S \bar{D}^{\dot{\alpha}}\bar{\S}
\right)=0 .
\ee
These equations can  
equivalently be expressed as
\be
\S = - \frac{1}{32\Lambda^4} \bar{D}_{\dot{\alpha}}\left(D^{\beta}\S D_{\beta}\S \bar{D}^{\dot{\alpha}}\bar{\S}
\right) + \bar{\Phi} 
\ee
where  $\Phi$ is a chiral superfield.
Hitting the above equation with $\bar{D}^2$ leads to a consistency condition
\be
\bar{D}^2 \bar{\Phi} = 0 ,
\ee
which implies that $\Phi$ is  a free chiral superfield. 
In fact, $\S$ can be written as 
\be
\S = H + \bar \Phi ,
\ee
where $H$ satisfies   the equations of motion

\be
\label{HHHH}
H = 
- \frac{1}{32\Lambda^4} \bar{D}_{\dot{\alpha}}\left(
D^{\beta}H D_{\beta}H \bar{D}^{\dot{\alpha}}\bar{H}
\right).
\ee
It is now straightforward to solve equation (\ref{HHHH}) in terms of  a constrained chiral superfield
subject to  (\ref{DDX}) and (\ref{XX}) by identifying $H$ (up to a phase)
with the  goldstino chiral superfield $X_{NL}$ 
\be
\label{HX}
 H=  X_{NL}\, .
\ee
Let us verify that (\ref{HX}) indeed solves (\ref{HHHH}).
From (\ref{XX}) one finds

\be
\label{XX1}
D^{\beta} X_{NL} D_{\beta} X_{NL} = - X_{NL} D^2 X_{NL},
\ee
whereas, (\ref{DDX}) gives
\be
\label{XX2}
X_{NL} \bar D^2 \bar X_{NL} &=& 4 \sqrt{2} \Lambda^2  X_{NL} ,
\\
\label{XX3}
X_{NL} D^2 X_{NL} &=& 4 \sqrt{2} \Lambda^2 X_{NL} + 8 X_{NL} \bar X_{NL} \bar \Psi .
\ee
For the right part of (\ref{HHHH}), by  using (\ref{HX}) we have
\be
\nn
& -& \frac{1}{32\Lambda^4} \bar{D}_{\dot{\alpha}}\left(
D^{\beta} X_{NL} D_{\beta} X_{NL} \bar{D}^{\dot{\alpha}}\bar X_{NL}
\right) 
\\
\nn
&=&
\frac{1}{32\Lambda^4} \bar{D}_{\dot{\alpha}}\left(
X_{NL} D^2 X_{NL} \bar{D}^{\dot{\alpha}}\bar X_{NL}
\right) 
\\
\nn
&=&
\frac{1}{32\Lambda^4} \bar{D}_{\dot{\alpha}}\left\{
\Big{(} 4 \sqrt{2} \Lambda^2 X_{NL} + 8 X_{NL} \bar X_{NL} \bar \Psi \Big{)} \bar{D}^{\dot{\alpha}}\bar X_{NL}
\right\} 
\\
\nn
&=&
\frac{1}{32\Lambda^4} \bar{D}_{\dot{\alpha}}\left\{
\Big{(} 4 \sqrt{2} \Lambda^2 X_{NL}  \Big{)} \bar{D}^{\dot{\alpha}}\bar X_{NL}
\right\} 
\\
\nn
&=&
\frac{1}{4 \sqrt{2} \Lambda^2} 
X_{NL} \bar{D}^2 \bar X_{NL} 
\\
\nn
&=&
X_{NL},
\ee
where we have used the identities (\ref{XX}), (\ref{XX1}), (\ref{XX2}) and (\ref{XX3}).
Thus, the equations of motion for the superfield $\S$ are solved in terms of a free chiral multiplet 
($D^2 \Phi = 0$),
and a constrained chiral superfield ($H= X_{NL}$).
Therefore, $\S$ describes on-shell a free chiral multiplet and a goldstino superfield.  
We should note however, that although (\ref{HX}) is a solution, we have not 
proven that it is unique.  

The component fields of the $\S$ multiplet can be deduced from the relation 
\be
\label{SXF}
\S = X_{NL} + \bar \Phi.
\ee
From eq.~(\ref{SXF}) the fields $F$ and $\l_\alpha$ of $\S$ are identified 
as the appropriate component fields of the constrained chiral superfield $X_{NL}$ since
\be
\l_\alpha = \frac{1}{\sqrt{2}} D_\alpha \S | =\frac{1}{\sqrt{2}} D_\alpha X_{NL}|
\ee
and
\be
F = - \frac{1}{4} D^2 \S | = - \frac{1}{4} D^2 X_{NL}| .
\ee 
Thus, we can deduce their equations of motion just from the $X_{NL}$.
On-shell we have
\be
X_{NL} = \frac{\l^2}{2 F} + \sqrt{2} \theta \l +\theta^2 F
\ee
with \cite{ks}
\be
\label{gF}
F &=& - \sqrt{2} \Lambda^2 
\left( 1 
+\frac{\bar \l^2}{16 \Lambda^8} \p^2 \l^2 
-\frac{3}{256 \Lambda^{16}} \l^2 \bar \l^2 \p^2 \l^2 \p^2 \bar \l^2
\right),
\\
\label{gold}
i \bar \s^{\mu \dot \alpha \alpha } \p_\mu \l_\alpha &=& \frac{1}{4 \Lambda^4} \bar \l^{\dot \alpha} \p^2 \l^2
-\frac{1}{64 \Lambda^{12}} \bar \l^{\dot \alpha} \l^2 \p^2 \l^2 \p^2 \bar \l^2
-\frac{1}{64 \Lambda^{12}}  \bar \l^{\dot \alpha}\p^2 (  \l^2  \bar \l^2\p^2 \l^2).
\ee
Equation (\ref{gold}) is the equation of motion for the goldstino 
and eq.(\ref{gF}) is the solution for $F$ in terms of the goldstino as anticipated.
From the chiral multiplet we can easily identify $\psi_\alpha$ as the fermion of the chiral multiplet $\Phi$, since
\be
\psi_\alpha =\frac{1}{\sqrt{2}} D_\alpha \bar \S | =\frac{1}{\sqrt{2}} D_\alpha \Phi | .
\ee
On-shell, $\Phi$ is a free chiral superfield so that  
\be
\Phi = A_\Phi + \sqrt{2} \theta \psi +\theta^2 F_\Phi 
\ee
with
\be
\p^2 A_\Phi &=&0
\\
 \bar \s^{\mu \dot \alpha \alpha } \p_\mu \psi_\alpha &=& 0 
\\
F_\Phi &=& 0.
\ee
Thus, $\psi_\alpha$ is a free massless fermion.
From (\ref{SXF}) we have, for the scalar component $A$ of $\S$ 
\be
A = \bar A_\Phi + \frac{\l^2}{2 F},
\ee
so that 
this component of $\Sigma$
is solved in terms of the free scalar of the chiral multiplet and  the goldstino.
The last two auxiliary fields $P_\mu$ and $\chi_\alpha$ can be specified similarly.
For the complex vector auxiliary  $P_\mu$ we have
\be
P_{\alpha \dot \alpha } 
= \bar D_{\dot \alpha} D_{\alpha} \S | 
= \bar D_{\dot \alpha} D_{\alpha} X_{NL} | 
= -2i \s^{\mu}_{\alpha \dot \alpha} \p_\mu \left( \frac{\l^2}{2 F}\right)
\ee
whereas for $\chi_\alpha$ we find
\be
\chi_{\alpha } 
= \frac{1}{2} \bar D_{\dot \alpha} D_{\alpha} \bar D^{\dot \alpha} \bar \S | 
= \frac{1}{2} \bar D_{\dot \alpha} D_{\alpha} \bar D^{\dot \alpha} \bar X_{NL} | 
=i \s^{\mu}_{\alpha \dot \alpha} \p_\mu \bar \l^{\dot \alpha}.
\ee

Such a model of SUSY breaking can be considered as a hidden sector. Then, couplings to the visible sector can be 
introduced through  the interactions
\be
{\cal L}_{\rm int}=-\frac{m_i^2}{2 \Lambda^4}\int d^4 \theta \S \bar{\S}\, \Phi^i\bar{\Phi}^i-
\frac{m_g}{4\Lambda^4}\int d^4 \theta \S \bar{\S}\, \Big{(}W^{\alpha}W_\alpha+\bar W_{\dot{\alpha}}\bar W^{\dot{\alpha}}\Big{)}
\ee
where $\Phi^i$ are chiral matter in the visible sector and $W_\alpha$ is the supersymmetric field strength of 
vectors. In the susy breaking vacuum,  $m_i,m_g$ are just 
soft masses for the  scalars of the chiral multiplets of the visible sector   and the gauginos, respectively.

\section{Conclusions}

It has been advocated in \cite{fer} that the addition of higher dimension operators to 
a supersymmetric theory may lead to the appearance of new vacua, 
where only one of them is continuously connected to the standard theory in the limit of removing
the higher dimension operators. This is possible, if the equations of motion for the auxiliaries 
have more than one solutions which satisfy the appropriate conditions. 
In \cite{fer}, some examples were discussed, none of which however realized that proposal. Here we have provided 
an example, where the proposal works. This is achieved by employing a complex linear multiplet, in which the 
quadratic term of its scalar  auxiliary fields has opposite sign of the corresponding term in a chiral multiplet action. Therefore,
by adding an appropriate ghost-free higher dimension operator, a potential is induced according to \cite{fer,ovrut,Far1,Far2}.
This potential, has a second non-supersymmetric vacuum at a non-zero value of the scalar auxiliary besides the 
supersymmetric one.  In the susy breaking vacuum,
the propagating fields are the scalar and the fermion of the complex linear multiplet and the goldstino mode of the 
broken supersymmetry. Interesting enough, the goldstino mode turns out to be one of the auxiliary fermions of the complex 
linear multiplet, which now propagates in the new non-supersymmetric vacuum.  The coupling of this model to 
supergravity is an interesting project that we leave for future work.

\section*{Acknowledgements}

We would like to thank Ignatios Antoniadis for discussions. 
This research was implemented under the “ARISTEIA” Action of the “Operational Programme Education and 
Lifelong Learning” and is co-funded by the European Social Fund (ESF) and National Resources. 
It is partially supported by European Union's Seventh
Framework Programme (FP7/2007-2013) under REA grant agreement n. 329083.
S.F. is supported by ERC Advanced Investigator Grant n. 226455 Supersymmetry, Quantum Gravity
and Gauge Fields (Superfields). M.P. is supported in part by NSF grant PHY-0758032. M.P. would
like to thank CERN for its kind hospitality and the ERC Advanced Investigator Grant n. 226455 for
support while at CERN.

\vskip.5in
\noindent
{\bf {APPENDIX}}
\vskip.3in

It should be noted that, instead of (\ref{ep1}), one could  
consider the following more general Lagrangian
\be
\label{ep11}
{\cal L}'_{EP}=\int d^4\theta \  \frac{1}{64}\, {\cal U}(\S,\bar{\S}) \   
D^{\alpha}\S D_{\alpha}\S \bar{D}_{\dot{\alpha}}\bar{\S}\bar{D}^{\dot{\alpha}}\bar{\S},
\ee
where, ${\cal U}$ is a real, strictly positive, but otherwise arbitrary function of $\S$ and $\bar \S$ with 
mass dimension $(-4)$. As we will see in the moment,  a potential emerges for 
the complex scalar $A$ of the complex linear multiplet 
$\S$. 
The component form of the bosonic part of  eq.~(\ref{ep11}) is 
\be
\label{ep21}
{\cal L}'^B_{EP}=\frac{1}{64}{\cal U}\ P^{\mu} P_{\mu} \bar{P}^{\nu} \bar{P}_{\nu} 
+\frac{1}{16}P_{\mu}\bar{P}^{\mu}{\cal U}F\bar{F}
+\frac{1}{4}\ {\cal U} F^2\bar{F}^2 ,
\ee
where ${\cal U}={\cal U}(A,\bar A)={\cal U}(\S,\bar{\S})\Big{|}$.
Then, the bosonic part of the  Lagrangian 
\begin{align}
{\cal{L}}'_\S&={\cal L}_0+{\cal L}'_{EP}\nonumber \\
&=\int d^4\theta \ \left(-\S\bar{\S}+ \frac{1}{64}\, {\cal U}(\S,\bar{\S}) \   
D^{\alpha}\S D_{\alpha}\S \bar{D}_{\dot{\alpha}}\bar{\S}\bar{D}^{\dot{\alpha}}\bar{\S}\right),  \label{Lo1}
\end{align}
is
\begin{align}
\nn
{\cal L}'^B=
&-F\bar{F}+A\p^2\bar{A}-\frac{1}{2}P_{\mu}\bar{P}^{\mu}
\\
\label{Lang21}
&+\frac{1}{64}{\cal U}\ P^{\mu} P_{\mu} \bar{P}^{\nu} \bar{P}_{\nu} 
+\frac{1}{16}P_{\mu}\bar{P}^{\mu}{\cal U}F\bar{F}
+\frac{1}{4}\ {\cal U} F^2\bar{F}^2.
\end{align}
From the equations of motion for the complex auxiliary vector we find again that 
\be
\label{P=00}
P_{\mu}=0,
\ee
whereas the equations of motion for the auxiliary scalar are now
\be
\label{FFf}
F\left(1-\frac{{\cal U}}{2}F\bar{F}\right)=0.
\ee
There are  again two solutions:
\be
&&(i) ~~~~F=0\, ,\label{vv}\\
&&(ii)~~F\bar F
=\frac{2}{{\cal U}(A,\bar{A})}  .\label{v222}
\ee
The first is the supersymmetric one while the second 
breaks supersymmetry.
Plugging back eqs.~(\ref{P=00}) and (\ref{FFf}) into (\ref{Lang21}) we find
\be
\label{Lang3}
{\cal L}^B=A\p^2\bar{A}-\frac{1}{{\cal U}(A,\bar{A})}.
\ee
We see now that a potential has emerged
\be \label{EP4}
{\cal V}_{EP}=\frac{1}{{\cal U}(A,\bar{A})}.
\ee
For example one can have 
\be
\label{U1}
{\cal U}(A,\bar{A})= \frac{1}{\Lambda^4 +m_A^2 A\bar{A}}
\ee
where $\Lambda$ is a mass scale. This case
leads to a scalar potential
\be \label{EP5}
{\cal V}=\Lambda^4 +m_A^2 A\bar{A}
\ee
i.e to a mass for the scalar $A$. 
The minimum of potential~(\ref{EP5}) is at $A=0$, which is  a supersymmetry breaking vacuum since 
\be
\label{susybreak}
<F\bar{F}> = 2 \Lambda^4\neq 0.
\ee
Another example is provided by 
\be
\label{U2}
{\cal U}(A,\bar{A})= \frac{1}{\Lambda^4 +\frac{\lambda}{4!}\big{(}A\bar{A}-\mu^2\big{)}^2},
\ee
which gives rise to a potential
\be
\label{V}
V=\Lambda^4 +\frac{\lambda}{4!}\big{(}A\bar{A}-\mu^2\big{)}^2.
\ee
In this case, the $U(1)$ global symmetry $A\to e^{i\alpha}A$ is broken at the vacuum $A\bar{A}=\mu^2$ 
where susy is also broken because
\be
<F\bar{F}> = 2 \Lambda^4\neq 0.
\ee

In general, the complex scalar multiplet can have  an arbitrary potential  in the susy breaking vacuum, specified 
by the arbitrary real positive function ${\cal U}(A,\bar A)$.

\end{document}